\documentclass[prd, aps, superscriptaddress, preprintnumbers, twocolumn, floatfix, nofootinbib]{revtex4-1}

\usepackage[utf8]{inputenc}
\usepackage[T1]{fontenc}

\usepackage{amsmath,amsfonts,amssymb}
\usepackage{physics}
\usepackage{mathtools}
\usepackage{bm}
\usepackage{cancel}

\usepackage{float}
\usepackage{wrapfig}

\usepackage[unicode=true,pdfusetitle,
 bookmarks=true,bookmarksnumbered=false,bookmarksopen=false,
 breaklinks=false,pdfborder={0 0 1},backref=false,colorlinks=true]
 {hyperref}
\hypersetup{
 linkcolor=blue, citecolor=cyan, urlcolor=red, filecolor=blue}

\usepackage{graphicx}

\numberwithin{equation}{section}

\newcommand{\da}{\downarrow}
\newcommand{\ua}{\uparrow}
\newcommand{\dm}[2]{\ket{#1}\!\bra{#2}}

\begin{document}

\title{On the Relative Distance of Entangled Systems in \\Emergent Spacetime Scenarios}

\author{Guilherme Franzmann}
\email{guilherme.franzmann@su.se}
\affiliation{Nordita, KTH Royal Institute of Technology and Stockholm University,\\
Hannes Alfvéns v\"ag 12, SE-106 91 Stockholm, Sweden}
\affiliation{Basic Research Community for Physics e.V., Mariannenstraße 89, Leipzig, Germany}

\author{Sebastian M. D. Jovancic}
\email{sebjov@kth.se}
\affiliation{KTH Royal Institute of Technology and Stockholm University,\\
Hannes Alfvéns v\"ag 12, SE-106 91 Stockholm, Sweden}

\author{Matthew Lawson}
\email{mmlawson@ucdavis.edu}
\affiliation{Savantic AB, Rosenlundsgatan 52, Stockholm, Sweden}

\date{\today}

\begin{abstract}

Spacetime emergence from entanglement proposes an alternative to quantizing gravity and typically derives a notion of distance based on the amount of mutual information shared across sub-systems. Albeit promising, this program still faces challenges to describe simple physical systems, such as a maximally entangled Bell pair that is taken apart while preserving its entanglement. We propose a solution to this problem: a reminder that quantum systems can have multiple sectors of independent degrees of freedom, and that each sector can be entangled. Thus, while one sector can decohere, and decrease the amount of total mutual information within the system, another sector, e.g. spin, can remain entangled. We illustrate this with a toy model, showing that only within the particles' momentum uncertainty there can be considerably more entanglement than in the spin sector for a single Bell pair. We finish by introducing some considerations about how spacetime could be tested in the lab in the future. 

\end{abstract}

\pacs{98.80.Cq}
\maketitle

\section{Introduction}

\begin{flushright}
\footnotesize
    \textit{``Of course spacetime \textbf{cannot} be emergent from entanglement, as we can place a Bell pair arbitrarily far apart from each other.''} \\ -- During a discussion session at  \href{https://qiss.uwo.ca/}{QISS '22 Conference}. 
\end{flushright}

A century has gone by since Planck discovered the quantum-mechanical nature of our Universe. Since then, the three fundamental interactions that mainly govern the microscopic scales have been quantized. Together, they compose the Standard Model of Particle Physics - the most accurate theory ever devised by us. And gravity does not fit in it. In spite of having its first mechanical description introduced by Newton centuries before, the program of quantizing gravity for arbitrarily high energies remains incomplete. 

Similarly to Yukawa's theory to describe the nuclear force between nucleons mediated by pions, quantum general relativity is a low-energy effective field theory \cite{Wallace:2021qyh}. As Yukawa's theory was superseded by quantum chromodynamics, where more fundamental degrees of freedom were introduced, the same is expected to happen with gravity. Thus, the quantized degrees of freedom in quantum general relativity, namely the spacetime fluctuations parametrized by the metric field, will no longer be fundamental in the final quantum gravity theory. 

Nonetheless, most attempts to reconcile gravity with quantum mechanics have insisted on keeping these degrees of freedom one way or another. String theory \cite{Polchinski:1998rq}, the most prevalent approach to quantum gravity, introduces other spacetime degrees of freedom that parametrize the strings' worldsheet. Meanwhile, some other approaches, such as loop quantum gravity \cite{Rovelli:2014ssa}, develop new ways of quantizing the same degrees of freedom from general relativity. Still, the single most significant insight since this program started came from Maldacena in 1997 \cite{Maldacena:1997re}. By introducing the anti-de Sitter/Conformal Field Theory (AdS/CFT) correspondence, he showed that a gravitational theory could be dual to a lower-dimensional quantum mechanical theory \textit{without} gravity. Since then, evidence about the emergent nature of spacetime has piled up. 

The research on spacetime emergence follows a long thread that started with findings by Bekenstein \cite{Bekenstein:1973ur}, who explored the thermodynamic properties of black holes and related the entropy of a black hole to its surface area. Later, Hawking \cite{Hawking:1975vcx} completed the thermodynamical description by showing that black holes indeed emit thermal radiation. Two decades later, the holographic principle was introduced by 't Hooft \cite{tHooft:1993dmi}, stating that the boundary of a bulk region of space encodes information about its interior. Meanwhile, Jacobson \cite{Jacobson:1995ab} showed that Einstein's equation of General Relativity could be seen as an equation of state resulting from the thermodynamic limit of local Rindler causal horizons. Despite these developments, theories without gravity \textit{ab initio} were still lacking until Maldacena \cite{Maldacena:1997re} proposed the AdS/CFT correspondence. It established a holographic relationship between AdS space and conformal field theories, also referred to as the gauge/gravity duality. Then, Ryu \& Takayanagi \cite{Ryu:2006bv} showed that in an AdS space there is a direct relationship between the entanglement entropy associated with bulk regions separated by a boundary surface where a conformal field theory is defined and the area of this boundary. Finally, van Raamsdonk \cite{VanRaamsdonk:2010pw} extended this relationship by suggesting that one could relate the boundary surface between bulk regions and the distance between them in an AdS/CFT setting.

\begin{figure}
\centering
 \includegraphics[scale=0.25]{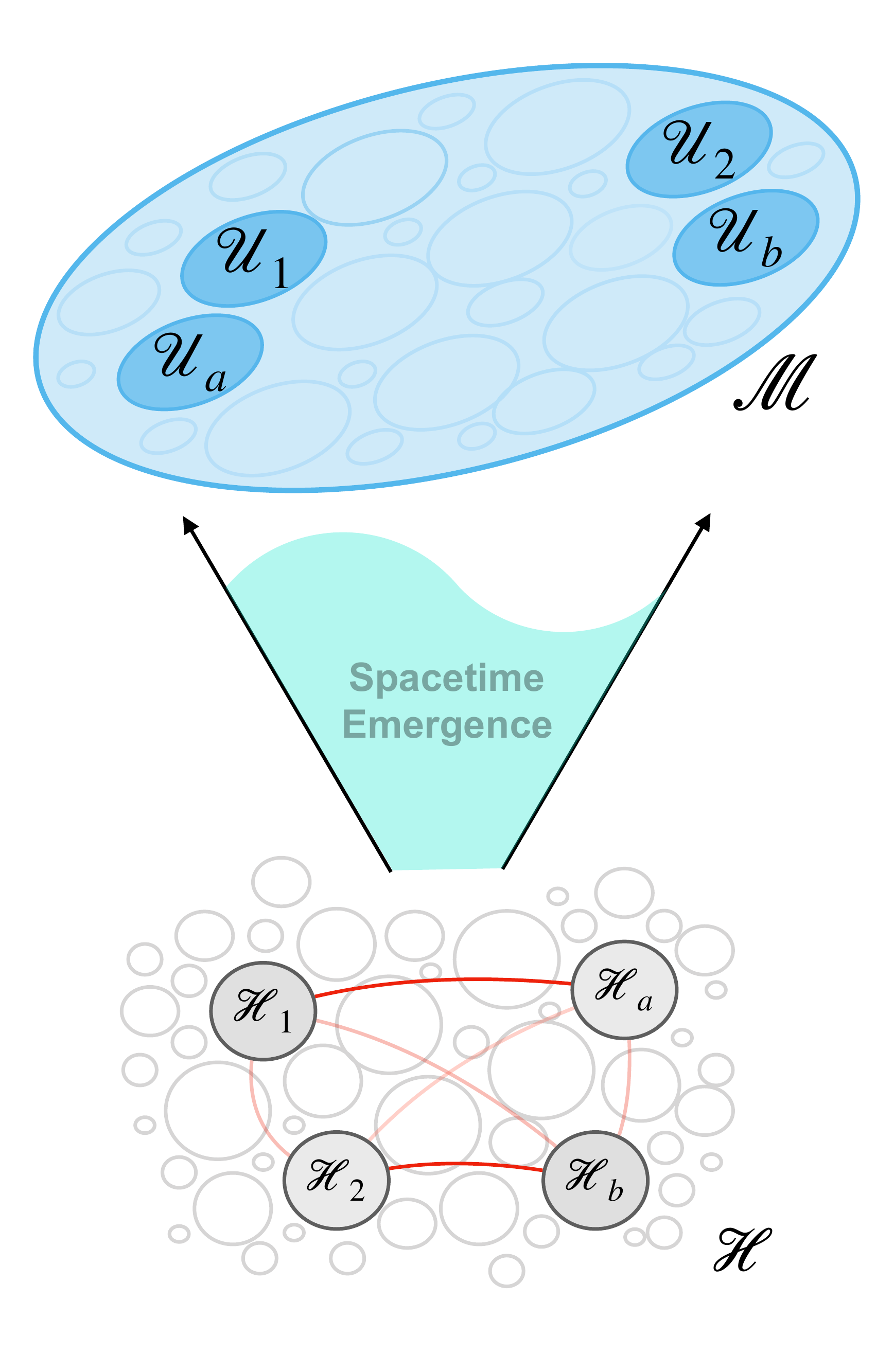} 
 \vspace{-0.5cm}
\caption{Representation of how space emerges from quantum mechanics. In the \textit{quantum} Hilbert space $\mathcal{H}$, different quantum sub-systems, $\mathcal{H}_p$, such that  $\mathcal{H} = \bigotimes_p \mathcal{H}_p$, are connected by their mutual information, which is larger for darker red lines. In the emergent \textit{classical} manifold $\mathcal{M}$, the relative distance of these sub-systems, $\mathcal{U}_p$, decays monotonically w.r.t. the amount of mutual information they share. }\label{fig:emergent}
\end{figure}

More recently, Cao et al. \cite{Cao:2016mst,Cao:2017hrv}
have combined much of this thread into a new research program.  It starts from a purely quantum-mechanical framework and its entanglement structure and derives classical spatial geometry satisfying Einstein's equation (as illustrated in Fig. \ref{fig:emergent}). In short, the program suggests that there is a mapping between the mutual information of quantum sub-systems and the classical geometry connecting them, giving rise to an emergent space purely defined in terms of the quantum information contained in the system. This research program builds up on several previous works \cite[e.g][]{Swingle:2009bg,Faulkner:2013ica,Czech:2015kbp,Faulkner:2017tkh,Raasakka:2017rmv}. Although this program still remains incomplete, we can argue that space, and perhaps time, will no longer be fundamental within this framework upon its completion. 

Nonetheless, even at this stage the program faces some challenges, as illustrated by the quote shown above. The mutual information of quantum sub-systems relates to the classical geometry connecting them such that the distance between sub-systems in the emergent geometry is a monotonically decreasing function of the mutual information (see Fig. \ref{fig:emergent}). The intuition here is that if quantum sub-systems are more entangled, then their locations in the emergent geometry will be closer; if they are not entangled, they will be as distant as allowed in the emergent space. The problem is that, for instance, a maximally entangled Bell pair will always have zero distance between its sub-components in the emergent geometry, even though we know that its sub-components can be arbitrarily separated. In fact, this resembles the ER = EPR conjecture introduced in \cite{Maldacena:2013xja}. The conjecture relates maximally entangled systems at the quantum level with wormhole-like geometries at the spacetime level. Thus, it is expected that despite the sub-components of a Bell pair be arbitrarily far away, they would still be connected by wormholes so that their distance is zero. Nevertheless, the conjecture is in tension with the fact that local observers do not see the wormhole geometry in their labs.

Thus, our objective is to show how one can recover the non-vanishing relative distances connecting the pair in the emergent space. We will show how the mutual information changes due to non-local entanglement perturbations and how that affects the emergent spacetime geometry. Hence, the typical notion of relative distances across entangled systems can be recovered instead of simply recovering wormholes as suggested in \cite{Maldacena:2013xja}. 

The paper is divided as follows: Sec. \ref{sec:spacetime_emerg} briefly reviews the spacetime emergence program introduced in \cite{Cao:2016mst,Cao:2017hrv}. In Sec. \ref{sec:closing}, we model a Bell state in ever-increasing levels of description and show one way to understand how its parts can be arbitrarily separated. Finally, in Sec. \ref{sec:lab} we discuss for the first time some ideas on how spacetime emergence can be tested in the lab, and then we conclude with discussion in Sec. \ref{sec:discussions}. 

\section{Spacetime Emergence and Relative Distances} \label{sec:spacetime_emerg}

The general strategy to reconstruct space from the Hilbert space introduced in \cite{Cao:2016mst,Cao:2017hrv} is: 
\begin{enumerate}
    \item Start with a Hilbert space  $\mathcal{H}$;
    
   \item Decompose the Hilbert space into a large numbers of factors  $\mathcal{H}=\bigotimes_p \mathcal{H}_p$ (see Appendix \ref{sec:TPS}). Every factor is finite-dimensional\footnote{In \cite{Bao:2017rnv,Raasakka:2019euf} is argued why we should expect that the Hilbert spaces associated with local regions of spacetime should be finite dimensional. Note that this implies a radical departure from QFT, where every point in spacetime gives rise to an infinite-dimensional Hilbert space for each field defined in it.};
   
   \item Consider only ``redundancy-constrained'' (RC) states, $|\psi\rangle_{\rm RC}$, that are a generalization of states in which the entropy of a region obeys an area law \cite{Wolf:2007tdq};
   
   \item Consider the weighted graph, $G$, with vertices represented by factors, $A_p,$ and edges represented by mutual information among them (see Appendix \ref{entropy}), $I(A_p\!:\!A_q) = S(A_p) + S(A_q) - S(A_p\!:\!A_q)$. Then, define a metric on the graph connecting the factors; 
   
   \item Construct a metric graph, $\Tilde{G}$, with smooth, flat geometries by mapping $G \rightarrow \Tilde{G}$;
   
    \item Consider perturbations on the quantum state, $|\delta \psi \rangle$, which deforms the entanglement structure of $\mathcal{H}$ and show that this yields local curvature proportional to the local change in entropy; 
    
    \item Finally, relate the change in entropy to the change of entropy of an effective field theory. Then, recover the Einstein's equation by means of the entanglement first law  $\delta S = \delta \langle K \rangle$ \cite{Blanco:2013joa,Faulkner:2013ica,Swingle:2009bg,Lin:2014hva},  relating the change in entropy to the change of the vacuum expectation value of the modular Hamiltonian, $K \equiv \ln \rho$, where $\rho$ is the density matrix of the unperturbed state.
\end{enumerate}

For the remainder of the paper, we will be mostly concerned in defining the metric in the emergent geometry between any two factors. So we now focus on items $3$ and $4$, and later on $5$ when we discuss how entanglement perturbations can disturb the line element between any two quantum systems. 

The overall expectation is that systems which are near in the emergent geometry share more mutual information than systems which are further apart \cite{VanRaamsdonk:2010pw}. To implement that, the metric graph $\Tilde{G}$ is assumed to have the same vertices as $G$ but with re-weighted edges by 
\begin{equation}
    w(A_p,A_q)=\begin{cases}
    l_{\rm RC} \Phi(I(A_p:A_q)/I_0),\quad &p\neq q\\
    0 &p=q \,,
    \end{cases}
\end{equation}
for some function of the mutual information $\Phi(I(A_p\!:\!A_q)/I_0)$ where $I_0= \max\{I(A_p\!:\!A_q)\}$, such that $\Phi$'s argument $\in [0,1]$, and $l_{\rm RC}$ is the scale of the RC states. The function $\Phi$ is defined to be a monotonically decreasing function such that $\lim_{x\rightarrow 1} \Phi(x) \rightarrow 0$ and  $\lim_{x\rightarrow 0} \Phi(x) \rightarrow \infty$\footnote{Naturally, the true form of this function will be more complicated than that if we expect to recover arbitrary spatial geometries}. A suitable candidate would be $\Phi(x) = -\log(x)$, which is expected in the ground state of a gapped system \cite{Wolf:2007tdq}. 

The vertices $p$ and $q$ are connected by a large set of vertices with many possible paths, such as $P=\left\{p_0=p, p_1,...,p_{k-1},p_k=q \right\}$, so the minimal path $P$ giving the distance function $d(A_p,A_q)$ is chosen to be the path which minimizes the sum of weights
\begin{equation}
    d(A_p,A_q) = \min_P \left \{ \sum_{n=0}^{k-1} w(p_n,p_{n+1}) \right \} \,,
\end{equation}
which by construction satisfies the properties of a metric, since the mutual information is symmetric, positive and a scalar, and the minimization of weights along a path satisfies the triangle inequality.

At this stage, we can contemplate our challenge: if we take a simple Bell pair and compute the mutual information between each qubit, the shortest path between them vanishes, as they are maximally entangled. One interpretation is to consider this as an implementation of ER = EPR, where the vanishing metric could be interpreted as describing the presence of a wormhole between $A_p$ and $A_q$. Nonetheless, we still need to understand how come we do not see the wormhole in a lab, as in the lab we can create entangled particles and separate them arbitrarily, in principle. This is the criticism against the program highlighted by the quote at the beginning of the paper. 

We aim to provide a way out of this conundrum. The basic idea is simple: when we talk about quantum systems, we  often only include some of their degrees of freedom, e.g. in the case of the Bell pair, we typically only model the pair by describing its spin state. We will see that by including other degrees of freedom, the Hilbert space gets extended by the presence of multiple sectors. While each sector can have a state which remains maximally entangled, the full state does not need to correspond to a maximally entangled one. Thus, the mutual information is not maximal and the distance between sub-systems can be different than zero. The null distance, or formation of wormholes between sub-systems, can be seen as an artifact of our ignorance of the full system.

\section{Closing down wormholes}\label{sec:closing}

We illustrate the discussions from last section by considering an initial particle of mass $M$ and spin $1$ that decays into two equal particles of mass $m$  and spin $\frac{1}{2}$. The pair is shared between Alice and Bob and we study it in ever-increasing levels of descriptions. For illustration purposes, we mostly focus on two qubits in an entangled triplet state for the spin sector, but all the considerations can be easily generalized. 

\subsection{Vanilla Bell pair}

Initially, we consider only the existence of the pair with the following TPS, $\mathcal{T}:\mathcal{H} \rightarrow \mathcal{H}_{\rm AB}=\mathcal{H}_{\rm A}\otimes\mathcal{H}_{\rm B}$, and we assume their state to be an entangled triplet qubit, 
\begin{equation} \label{eq:AB_1}
\small
    \ket{\psi}_{\rm AB} = \frac{1}{\sqrt{2}}\left(\ket{\uparrow}_{\rm A}\otimes\ket{\uparrow}_{\rm B} + \ket{\da}_{\rm A}\otimes\ket{\da}_{\rm B}\right) \equiv \frac{1}{\sqrt{2}}\Big(\ket{\uparrow\uparrow}+\ket{\da\da} \Big) \,.
\end{equation}
\normalsize
We can easily see that Bob's and Alice's reduced density matrices are maximally mixed and that the von Neumann entropy associated with them is $S(\rho_i)= \log 2$, where $i=\{A,B\}$, while their mutual information is $I(A\!:\!B)= 2\log 2$. Thus, the state saturates their mutual information, which is upper bounded by the sum of the logarithm of each Hilbert spaces' dimensions (see Appendix \ref{entropy}). This can be easily generalized by considering two qudits instead, such that $S(\rho_i)= \log N$ and $I(A\!:\!B)= 2\log N$, where $N$ is the dimension of each Hilbert space factor, $\mathcal{H}_i$.

Now we consider the presence of a simple environment surrounding Alice made of a simple qubit. The extended Hilbert space reads, $\mathcal{H} \rightarrow \mathcal{H}_{\rm A}\otimes\mathcal{H}_{\rm B} \otimes \mathcal{H}_{\rm E}$, and after allowing Alice's particle to entangle with the environment by some unitary transformation, $U$, the final state is
\begin{equation}\label{eq:AB_2}
    \ket{\psi}_{\rm AB}\otimes\ket{e} \xrightarrow{\text{U}} \ket{\psi'}=\frac{1}{\sqrt{2}}\big(\ket{\uparrow\uparrow\uparrow} + \ket{\da\da\da} \big) \,.
\end{equation}
Different than when the pair is isolated, now the density matrix associated with Alice's and Bob's pair does not correspond anymore to a pure state, and $S(\rho_{\rm AB}) = \log 2$, while the entropy associated with each subsystem remains the same, $S(\rho_i)= \log 2$. This implies that the mutual information between Alice and Bob is now $I(A\!:\!B)= \log 2$, and does not saturate the upper bound anymore. This is related to the monogamy of entanglement \cite{Witten:2018zva}, which prevents two qubits from being maximally entangled if any of them is also entangled with another qubit.

As we just saw, $I_{\ket{\psi}}(A\!:\!B) > I_{\ket{\psi'}}(A\!:\!B)$, while $I_{\rm max}(A\!:\!B)$ does not depend on the state but only on the dimensions of $\mathcal{H}_i$. The emergent metric built in Sec. \ref{sec:spacetime_emerg} assigns zero distance to maximally entangled sub-systems. That was the case for the Bell pair without the presence of the environmental qubit, but it is no longer the case when the pair gets entangled with it.  In fact, using the $\Phi(x) = - \log x$ \cite{Cao:2016mst,Qi:2013caa,VanRaamsdonk:2010pw},  we have 
\begin{subequations}
\begin{align}
  &  w(A\!:\!B)_{\ket{\psi}} = -l_{RC} \log 1 = 0 \,,
\\
  &  w(A\!:\!B)_{\ket{\psi'}} = l_{RC} \log 2 \,.
\end{align}
\end{subequations}
Thus, at this level, we can start to understand how the relative distance between the sub-systems can be different than zero. Nonetheless, we are still not done, as the $\ket{\psi'}$ does not correspond any longer to a maximally entangled spin state, so it could not possibly model a Bell pair that has its sub-systems arbitrarily far apart. 

\subsection{What do quantum states describe?}

The typical Bell states representing entangled spin systems should not be confused with an exhaustive description of these systems. There are many other quantum degrees of freedom that are tacitly suppressed when we simply model the spin sector. In fact, the typical example of a decaying particle into two particles that have their spin states entangled should also conserve their momentum, for example. In the rest mass frame,  the state \eqref{eq:AB_1} would be generalized to
\begin{equation} \label{eq:AB_3}
    \ket{\psi} = \frac{1}{\sqrt{2}} (\ket{\ua\ua} + \ket{\da\da}) \otimes \ket{p,-p}\,,
\end{equation}
where the Hilbert space is extended, $\mathcal{H} = \mathcal{H}_{\rm A}\otimes \mathcal{H}_{\rm B} = \mathcal{H}_{\rm A}^{\rm s} \otimes \mathcal{H}_{\rm B}^{\rm s} \otimes \mathcal{H}_{\rm A}^{\rm p} \otimes \mathcal{H}_{\rm B}^{\rm p}$, to also take into account the momentum degrees of freedom, here simply represented as a two-level state (we develop more on this below) as $p$ can be computed classically in terms of $m$ and $M$ and each particle has either $p$ or $-p$ as its momentum in the rest frame of the initial system. In other words, we consider two particles with conserved momentum and spin. In the non-relativistic limit, the spin and momentum sector lead to two independent sectors in the Hilbert space and we can expect that each sector has separable contributions to the overall density matrix, such that 
\begin{equation}
    \rho_{\rm AB} = \rho^{\rm s}_{\rm AB}\otimes\rho^{\rm p}_{\rm AB} \,.
\end{equation}
Still, each sector can give rise to respective degrees of freedom that are entangled, 
\begin{equation}
    \rho^{\rm s}_{\rm AB} \neq \rho^{\rm s}_{\rm A}\otimes\rho^{\rm s}_{\rm B},\quad\quad \rho^{\rm p}_{\rm AB} \neq \rho^{\rm p}_{\rm A}\otimes\rho^{\rm p}_{\rm B} \,.
\end{equation}
Then, the total mutual information is 
\begin{align}
    I(A:B) &= S(\rho_{\rm A}) + S(\rho_{\rm B}) - S(\rho_{\rm AB}) \nonumber \\
    &= I(A^s:B^s) + I(A^p:B^p)\,,
\end{align}
which trivially implies $I(A:B)\geq I(A^s:B^s)$. 

Therefore, we can have an entangled Bell pair whose total mutual information exceeds the spin sector's mutual information. Qualitatively, we consider that initially an entangled Bell pair has maximal mutual information both in spin and momentum sector, and as we separate them, the mutual information in the momentum sector decreases. Naturally, when the size of the Hilbert space for the momentum sector far exceeds the size of the Hilbert space for the spin sector, the mutual information of the spin sector is negligible in comparison and so
\begin{equation}\begin{split}
    I(A\!:\!B) \approx I(A^p\!:B^p)\,,
\end{split}\end{equation}
and correspondingly the line element of a Bell pair would be largely independent of the spin correlation.

Let's see how we can think about this for the momentum sector now.

\subsection{Realistic Bell pair}



We  now consider a more realistic example of a Bell pair with both a spin and a momentum sector. Once again we consider the decay of a particle $M\rightarrow m + m$ and suppose the momentum of the initial particle is $\Delta p$ in its frame due its the momentum uncertainty. Since we expect we can localize our particles in our labs to measure their spin, that naturally constrains the uncertainty in the position by some characteristic scale of our apparatus, $l_{ \rm app}$. By saturating the uncertainty principle, then the momentum uncertainty is upper bounded by $ p_{\rm app} \sim \hbar/l_{\rm app}$\footnote{In general, the upper bound can be much higher, $\Delta p_{\rm max} \sim  mc$, which would correspond to having enough energy to create another particle of the same mass. Considering this momentum uncertainty, and saturating the uncertainty principle, one can find a minimum uncertainty for the particle's location that is half of its reduced Compton wavelength, $\Delta x \geq \hbar / (2mc)$. 
}.

As we briefly discussed in Sec. \ref{sec:spacetime_emerg}, we only consider finite-dimensional Hilbert spaces. As a consequence, this naturally regularizes the entropy and the mutual information, as these quantities typically diverge in field theories (see \cite{Bianchi:2019pvv} for comments and alternatives proposals). Naturally this poses a problem for the position and momentum operators, that can only satisfy the Heisenberg algebra when defined in infinite-dimensional Hilbert spaces. However, this can be circumvented by utilizing Generalized Pauli Operators (GPO) to represent the finite-dimensional analog of the position and momentum operators, which then satisfies the Weyl's exponential form of the Heisenberg algebra \cite{Singh:2018qzk}. By taking the infinite limit of the the Hilbert space's dimension, these operators do satisfy the Heisenberg algebra. 

Notice that by imposing an upper bound on the momentum uncertainty does not necessarily lead to having a finite-dimensional Hilbert space. Thus, we further assume that the momentum must be lower bounded by some infrared (IR) scale, $l_{\rm IR}$, and its eigenvalues are discretized in terms of it. We postulate it to be on the scale of the cosmological constant \cite{Yargic:2019lga}, $\Lambda$, such that $l_{\rm IR} \sim \Lambda^{-\frac{1}{2}}\sim 10^{26} \, {\rm m}$ . We suppose therefore that the lower bound is given by $p_{\rm IR}= \hbar \Lambda^{\frac{1}{2}}$ which is on the order $10^{-60} \, {\rm kg\cdot m/s}$. Thus, we bound the uncertainty in the particle's momentum by
\begin{equation}
    p_{\rm IR} \leq \Delta p \leq p_{\rm app}\,, \label{eq:bound}
\end{equation}
which applies to any of the particles being considered. 

After the initial particle decays, by momentum conservation we have
\begin{equation}
    \Delta p = p_1 + \Delta p_1 + p_2 + \Delta p_2\,, 
\end{equation}
which generalizes the conservation law used in Eq. \eqref{eq:AB_3}. Nonetheless, as the the overall momenta should still conserve, $p_1 = -p_2$ (alternatively, within the bound \eqref{eq:bound} typically $p_i/\Delta p_i \gg 1$), then we have
\begin{equation}
    \Delta p_1 +  \Delta p_2 = \Delta p\,.
\end{equation}
As we are in the rest frame of the initial particle, we are minimizing the uncertainty in position for $M$ and so the uncertainty $\Delta p$ associated with $M$ must be maximal, so $\Delta p= p_{\rm app}$.

Now we would like to model the momentum modes associated with the momentum uncertainties of these particles such that momentum is always conserved. We express the overall momentum state as a linear combination of all possible momentum states w.r.t. all possible uncertainties, summing from $p_{\rm IR}$ to $p_{\rm app}$ in steps of of $p_{\rm IR}$
\begin{equation}
    \ket{\pi} = \sum_{\Delta p_1=p_{\rm IR}}^{p_{\rm app} - p_{\rm IR}} \alpha_{\Delta p_1} \ket{p_1 + \Delta p_1, p_{\rm app}-(p_1 + \Delta p_1)}\,,
\end{equation}
where $\alpha_{\Delta p_1}$ determines the probability distribution of the states. This is an entangled state between the uncertainties around the momentum of each particle. Note that $p_{\rm app}/p_{\rm IR} \sim 10^{29}$ for a milimeter-size apparatus. Thus, the difference between them is large enough that we can consider that $p_{\rm app}$ is effectively an integer multiple of $p_{\rm IR}$. 

Since the momentum state only depends on the uncertainty, we can omit $p_1$, and by defining $N=p_{\rm app}/p_{\rm IR}$, and redefining the sum and its coefficients, the state can be written as 
\begin{equation}
\ket{\Delta \pi} =\sum_{n=1}^{N-1} \alpha_n \ket{n,N-n}\,,    
\end{equation}
where $\min \{N-1\} = 2$, corresponding to the entangled state $\ket{\Delta \pi} \varpropto \ket{p_{\rm IR},p_{\rm lab} - p_{\rm IR}} + \ket{p_{\rm lab} \sim p_{\rm IR},p_{\rm IR}}$, and $\max \{N-1\} \sim l_{\rm IR} mc/\hbar \, \textrm{(e.g. for an electron} \sim 10^{38}$), corresponding to $p_{\rm app}$ approaching the Compton's momentum of the particle, i.e. localizing it the most. The density matrix associated with the momentum sector is
\begin{equation}
        \rho^{\rm p} = \sum_{n,m=1}^{N-1} \alpha_n \alpha_m^* \dm{n,N-n}{m,N -m}\,,
\end{equation}
such that the reduced density matrix is 
\begin{equation}
        \rho^{\rm p} =\sum_{n=1}^{N-1} |\alpha_n|^2 \dm{n}{n}\,,
\end{equation}
and the mutual information in the momentum sector is given by
\begin{equation}
    I(A^{\rm p}\!:\!B^{\rm p}) = -2\sum_{n=1}^{N-1} |\alpha_n|^2 \log |\alpha_n|^2\,.
\end{equation}

To be quantitative, let's consider a maximally entangled state for simplicity, whereas a more realistic description would involve a non-flat probability distribution. Thus, we take $|\alpha_i|^2=\frac{1}{N-1}$, and the total mutual information contained alone in the momentum sector is 
\begin{equation}
    I(A^{\rm p}\!:\!B^{\rm p}) = 2\log (N-1).
\end{equation}
When the Bell pair is separated, it is this initial mutual information in the momentum sector that decreases as the momentum degrees of freedom decohere w.r.t. the environment. The more the initial system is localized, the more states are initially entangled (up to $N \sim l_{\rm IR} Mc/\hbar$). As the pair is more and more localized in Alice's and Bob's labs (see Fig. \ref{fig:rel_dis}), which corresponds to non-local entanglement perturbations of its initial state (see Appendix \ref{sub:entang_pert}), more of these modes are decohering, until all the mutual information in the momentum sector is depleted, dropping to zero once any of the systems is completely localized in any of the labs. The decoherence happens gradually, from the most IR modes ($\sim \Lambda_{\rm IR}$), thermal states in the environment that get entangled with the pair, to modes that typically further localize the particles in their labs ($\sim l_{\rm lab}$), all the way to the most UV ones ($\sim l_{\rm app}$), represented by Alice's and Bob's apparatus. Thus, as more momentum modes get entangled with the environment, more distance is allowed between the sub-systems. All along, we can preserve the spin sector untouched, remaining maximally entangled.

\begin{figure}
\centering
 \includegraphics[scale=0.23]{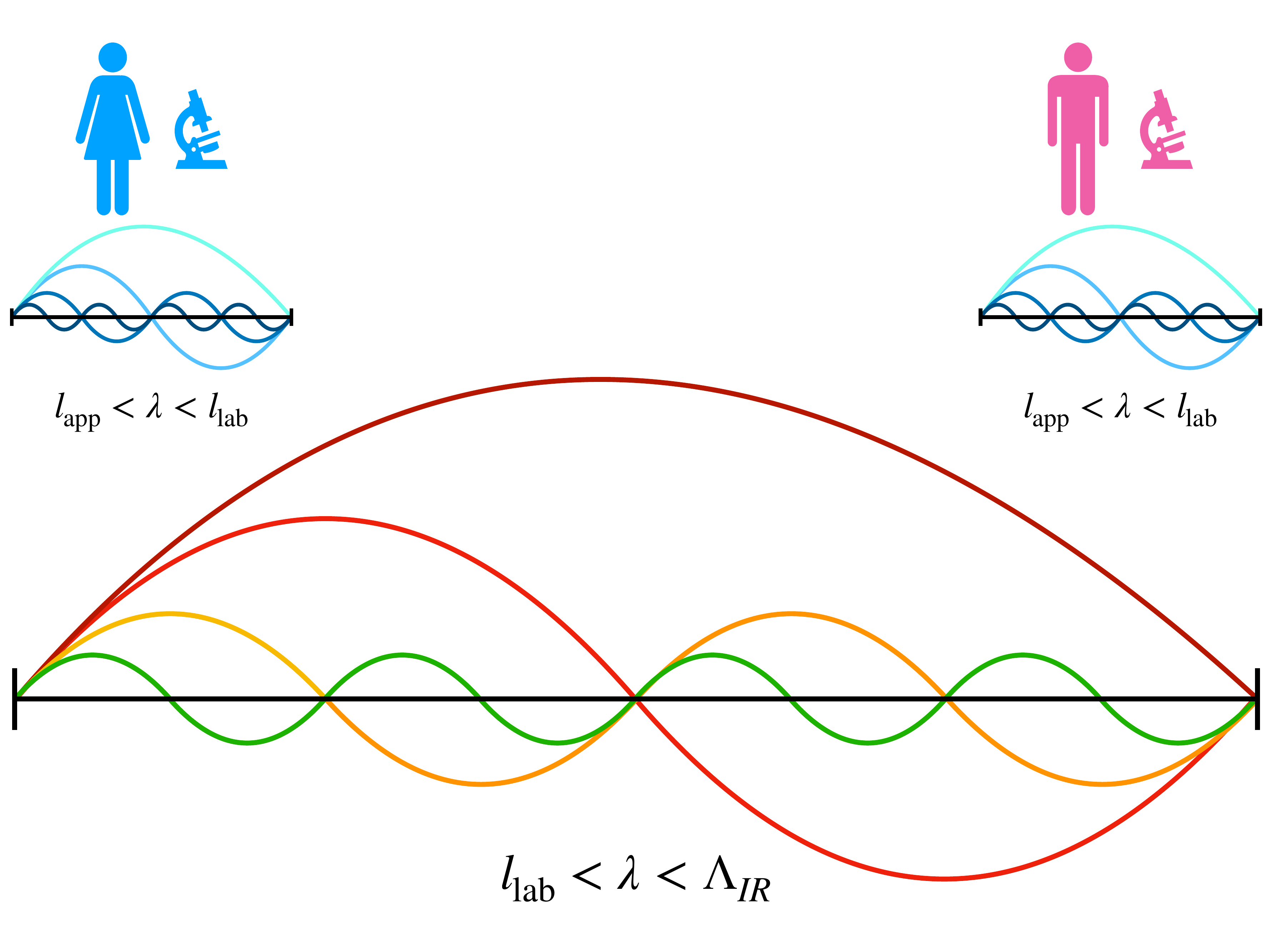} 
 \vspace{-0.5cm}
\caption{Different sets of modes associated with the system need to decohere so that local observers can infer the sub-systems' relative distance: i) $l_{\rm lab}<\lambda<\Lambda_{\rm IR} $ are the modes that decohere giving rise to the relative distance between the sub-systems, while ii) at each lab, $l_{\rm app}<\lambda<l_{\rm lab}$ are the modes that decohere with the local environment/apparatus, allowing the local observers to localize each sub-system.  Naturally, as more modes decohere with the environment, the overall mutual information across the sub-systems further decreases, therefore increasing their relative distance.}
\label{fig:rel_dis}
\end{figure}

\section{Spacetime emergence in the lab?}\label{sec:lab}

In recent years, applied methods of quantum information to gravitational-quantum systems led to the development of table-top experimental proposals that intend to exhibit the quantum nature of gravity. By demonstrating that gravity can induce entanglement, they will allow us to infer the existence of superimposed metric fluctuations \cite{Adlam:2022qzg}. For instance, in the Bose-Marletto-Vedral experiment \cite{Bose:2017nin,Marletto:2017kzi}, two particles start off in a superposition of two different spatial positions, leading to four different branches of the wavefunction. In each branch the gravitational interaction between the particles yields a different phase shift, entangling them. It is argued that if gravity can entangle two systems then we should conclude that is indeed quantum \cite{Marletto:2017kzi} (for a critical analysis of these proposals, see \cite{Anastopoulos:2022vvb}).

However, suppose space(-time) is emergent from more fundamental
quantum degrees of freedom. In that case, it is unclear what to
conclude from these experiments, as we do expect that these more
fundamental degrees of freedom are ruled by interactions that produce
entanglement, despite not being gravitational. One way to parse out
this degeneracy about the nature of gravity
is by probing its emergent nature in the lab. Here we discuss some of the
theory behind possible experimental setups and sketch possible
experimental prototypes.

\paragraph*{Non-local entanglement perturbations.} In the context of space emerging from the entanglement structure contained in the Hilbert space, we have seen that the mutual information across quantum sub-systems may gives rise to their emergent relative distance. One way of changing the amount of mutual information between sub-systems is to consider entanglement perturbations (see Appendix \ref{sub:entang_pert}), which will lead to fluctuations in the line element associated with these sub-systems in the emergent space \cite{Cao:2017hrv}. 

There is a way to induce such spacetime fluctuations. As the mutual information provides an upper bound on correlations in a system \cite{Wolf:2007tdq,VanRaamsdonk:2010pw},
\begin{equation}
I(C\!:\!D) \geq \frac{\left( \langle \mathcal{O}_C \mathcal{O}_D \rangle - \langle \mathcal{O}_C \rangle \langle \mathcal{O}_D \rangle \right)^2}{2 |\mathcal{O}_C|^2 |\mathcal{O}_D|^2} \,,
\end{equation}
we can consider varying any sort of correlations across two sub-systems, hence inducing a variation of their mutual information, and consequently  varying their emergent relative distance. Since we could modulate the mutual information very precisely at a set
frequency, a lock-in amplification type technique coupled with precise distance measurements could be used to
search for the corresponding distance modulation.

One natural system for realizing this type of measurement would be a
collection of hyperpolarized spins.
Large volumes of hyperpolarized atomic spins can be produced by
various techniques including ParaHydrogen-Induced
Polarization (PHIP), Signal Amplification by Reversible Exchange
(SABRE), or xenon hyperpolarization by
spin exchange with optically pumped rubidium atoms. All of these
techniques can produce on order liters
of liquid targets containing atomic spin polarization fractions of
order one. Using standard NMR pulse sequences
such as Carr-Purcell-Meiboom-Gill sequence one can modulate the spin
polarization direction of the spin ensemble at
regular intervals. With two such spin ensembles at opposite ends of a
very sensitive interferometer, one could
look for small fluctuations in the distance between the spin ensembles
at the modulation frequency.
While the correlation between the spin ensembles in this case would be
classical, this does not necessarily present a problem for the scheme.
For a classical mixture of separable states, represented by
$\rho_{\rm AB}^M=1/2 (\ket{\da\da}\bra{\da\da} +
\ket{\ua\ua}\bra{\ua\ua})$, $I(A\!:\!B)=\log 2$, half of the mutual
information corresponding to a maximally entangled qubit. Thus, the
classical version\footnote{Considering that fundamentally the world is quantum, we can imagine that all classical correlations are quantum in origin by the Schrödinger–HJW theorem, such that any mixed state can be represented as the partial trace of a pure state defined in an enlarged Hilbert space. Thus, manipulating seemingly classical correlations is another way of changing the entanglement across systems.} of the experiment will have less resolving power per
unit of correlation, but could have more total resolving power given
the sheer number of correlated spins. 

A quantum version of such an experiment could be a modification of
certain current atomic interferometry
initiatives. A typical atomic interferometer uses a highly correlated
ensemble of cold atoms (often a Bose-Einstein Condensate), launches it
on a geodesic trajectory, and uses a series of laser pulses to serve
as beam splitters for the matter wave. The spatially separated wave
packets then probe different regions of space before being interfered
with each other and the interference pattern measured. The spatially
separated wave packets naturally have a high degree of quantum
correlation, being part of the same quantum state. The spatial
separation allows the interferometer to very precisely measure the
separation of its own wavepackets by for example placing a gradient
field on the interferometer.

One could also imagine using a device such as the under-construction
MAGIS-100 interferometer \cite{Abe_2021} which is an
atomic interferometer designed to measure gravity waves. The MAGIS-100
interferometer uses two independently produced simultaneous populations
of cold atoms to do two independent simultaneous measurements to
detect
gravitational waves. Were the two populations produced in an entangled
state and the result compared to an identical experiment with
unentangled cold atom packets, the result would be an extremely
sensitive test of emergent spacetime
in this framework. This presents a significant experimental challenge
but is in principle possible \cite{doi:10.1126/science.aao2035}.

Classical and quantum versions of these proposals are currently being developed \cite{spacetime:emerglab}. 

\section{Discussion}\label{sec:discussions}

We have proposed one way to understand how, if relative distance between quantum sub-systems are to be understood as emerging from their shared mutual information, we can have maximally entangled systems which have their sub-components taken apart while preserving their entanglement.  Our proposal relies on the fact that our typical representation of quantum states oversees most of their degrees of freedom. Thus, it is expected that the full quantum state of a Bell pair, considered our toy model, lives in an extended Hilbert space that is considerably larger than just the two-dimensional Hilbert space used to model the spin sector. By decohering the extra sector associated with these other degrees of freedom, the mutual information shared across the sub-systems decreases considerably, allowing their relative distance to be different than zero.

To implement such an idea, in the Bell pair example we considered that the extended Hilbert space had to do with momentum modes. Is that reasonable if we are attempting to discuss space(-time) emergence? This is not unprecedented, as the idea that we live in a spacetime is actually constructed by inference from our measurements of momenta and energy. The notion that we all live in the same spacetime, a.k.a. as absolute locality, can be relaxed in the framework of relative locality \cite{Amelino-Camelia:2011lvm}. As we have not tried to address the absolute existence of spacetime, but rather just understand how maximally entangled systems can have non-vanishing relative distances in an already existing spacetime, we tacitly assumed a limit in which degrees of freedom associated with momentum are already present. This is intrinsically related to recovering quasi-classical dynamics, since we can find a TPS in which the quantum system is best represented by emerging degrees of freedom, such as the ones described by position and momentum, in which its Hamiltonian is local \cite{Carroll:2020gme}.

Other ideas have been proposed to explain how come excited maximally entangled states corresponding to composite systems can be taken apart. In \cite{Carroll_22,Cao:2017hrv}, they focus on the entanglement contained exclusively in the vacuum state surrounding excited quantum systems, which is believed to overwhelm the entanglement contained exclusively within the systems. We believe that both approaches are complementary.

As we start to further understand how entanglement can give rise to spatial distances, we also realize that the reconstruction remains incomplete. In particular, we still do not know how to precisely relate both as the map $\Phi(I(A\!:\!B))$ remains unknown. Nonetheless, we can strip down most of the proposal and keep only its core principle that directly relates entanglement and relative distances, allowing us to envision how such ideas can be tested in the lab in the future. We briefly sketched two proposals above and plan to elaborate on them in future works \cite{spacetime:emerglab}. 

Finally, we point out that a crucial assumption throughout the paper was that we had a particular tensor product structure defined in the Hilbert space. As the entanglement structure depends on the TPS of the Hilbert space, different TPS's will result in having different emergent geometries. Thus, we still need to clarify why it seems that there is a preferred emergent geometry in our world. It is conceivable that this puzzle is closely related to the transition quantum-classical \cite{Carroll:2020gme} and the imposition of locality \cite{Cotler:2017abq}. We will elaborate on these ideas in future works 
\cite{spacetime:emergTPS}. 

\newpage

\section{Acknowledgments} 
The work of G. Franzmann was supported by the Göran Gustafsson Foundation for Research in Natural Sciences and Medicine. Nordita is partially supported by Nordforsk. 

\appendix

\section{}

\subsection{Tensor Product Structures}\label{sec:TPS}
A Tensor Product Structure (TPS) $\mathcal{T}$ of a Hilbert space $\mathcal{H}$ is an equivalence class of isomorphisms 
\begin{equation}
    T:\mathcal{H} \rightarrow \bigotimes_p \mathcal{H}_p\,,
\end{equation}
where $T_1\sim T_2$ whenever $T_1 T_2^{-1}$ can be written as a product of local unitaries $\bigotimes U_p$ and permutations of sub-systems \cite{Cotler:2017abq}. In this regard, operators acting only on their respective factors of the Hilbert space introduces the notion of locality. 

Note that the entanglement structure of the Hilbert space is dependent on the choice of TPS.

\subsection{von Neumann Entropy and Mutual Information }\label{entropy}
Given a density matrix $\rho$, the entropy associated with it is given by the von Neumann entropy
\begin{equation}
    S = -tr(\rho \log \rho)\,.
\end{equation}
Since we can always diagonalize $\rho$, we can rewrite it in terms of its eigenvalues, $\lambda_i$, as
\begin{equation}\begin{split}
    S =-\sum_i \lambda_i\log \lambda_i \,.
\end{split}\end{equation}
We can see that the entropy associated with a pure state is zero, while for a maximally mixed state of dimension $N$, the entropy is maximal, $\log N$. The joint entropy for some partitioned Hilbert space for which the density matrix can be written as $ \rho = \bigotimes_p \rho_p$, is $S(\rho) = \sum_p S(\rho_p)$.

The mutual information between two sub-systems $A$ and $B$ is 
\begin{equation}
    I(A\!:\!B) = S(A)+ S(B) - S(A,B) \,,
\end{equation}
with the following properties \cite{Witten:2018zva}: i) positive: $I(A\!:\!B) \geq 0$; ii) upper-bounded: $I(A\!:\!B) \leq \log \, (\textrm{dim}(A))+ \log \, (\textrm{dim}(B))$; iii) symmetric: $I(A\!:\!B) = I(B\!:\!A)$, iv) $I(A\!:\!BC)\geq I(A\!:\!B)$.

\subsection{Entanglement Perturbations} \label{sub:entang_pert}

We can consider a variety of entanglement perturbations that change the amount of entanglement shared between Alice and Bob \cite{Cao:2016mst}. \textit{Local perturbations} are generated by some unitary operator, $U_{\rm AB}$, acting on the original system, $\mathcal{H}_{\rm AB}$. As the total entropy of the system does not change, the mutual information varies as $\delta I(A\!:\!B) = 2 \delta S_{\rm A}$. Meanwhile, \textit{non-local perturbations} entangle only part of the original system with new degrees of freedom, e.g. a local environment $\mathcal{H}_{\rm E}$. The non-local perturbation is instantiated by a unitary over the extended Hilbert space, $U_{\rm ABE} = U_{\rm AE}\otimes I_{\rm B}$, here acting over Alice's and environmental degrees of freedom. This can also be modeled as non-unitary transformations acting only on $\mathcal{H}_{\rm A}$. Due to monogamy of entanglement, non-local perturbations always lead to $\delta I (A\!:\!B) \leq 0$. This is also known as \emph{data processing inequality} \cite{Pawowski2009}.

\phantomsection
\addcontentsline{toc}{section}{References}

\let\oldbibliography\thebibliography
\renewcommand{\thebibliography}[1]{
  \oldbibliography{#1}
  \setlength{\parskip}{0pt}
  \setlength{\itemsep}{0pt} 
  \footnotesize 
}

\bibliographystyle{JHEP2}
\bibliography{References}

\end{document}